# The method of partial wave functions and the structure of time interval between the subsequent quantum transitions [1]

## A.G. Shkorbatov

*Institute for Low Temperature Physics and Engineering, National Ukrainian Academy of Sciences, Lenin Ave. 47, 61103 Kharkov, Ukraine*

We present the formulation of non relativistic quantum mechanics in the extended space $(u,x,t)$ where x and t are coordinates of particles and time, and u – an additional real parameter that corresponds to generalized virial - an integral over path on potential energy of the particle. If the u value is defined, the state of a particle is described by the partial wave function $F(u,x,t)$ (PWF). The wave function is obtained from $F(u,x,t)$ by special integral transform. The "drift" wave equation (DWE) for $F(u,x,t)$ is introduced. The integral invariants of $F(u,x,t)$ evolution are obtained. The sets of orthonormal solutions for stationary DWE are obtained; the moments of distribution of are calculated. The process of "aging" of PWF is described. Effective time interval necessary for the creation of the state is calculated within the limits of PWF formalism. The structure of the time interval between the processes of consequent quantum transition is investigated. The theoretical basis for the describing counterfactual situations in quantum mechanics is examined.

# I. INTRODUCTION

The structure of Feynman continual integral for wave function of the non-relativist particle without spin is analyzed in the present work. Compared to the powerful apparatus of quantum electrodynamics and grandiose constructions of supersymmetry and superstring theories, the basic for our analysis Schrödinger equation looks as an archaic instrument of theoretical physics. Nevertheless, from the mathematician's point of view "Schrödinger equation is unlimited, as mathematics itself" [1]. Physicists often address this equation because the combination of its dynamic form and the statistical interpretation of its solutions. The attempts to "improve" the dynamic scheme of quantum mechanics by the addition of "hidden" parameters (see review [2]), detailing the measurement process [3], introduction of "measurement operators" [4], and stochastic additions to Hamiltonian, are numerous. Regretfully, according to the old analysis by von Neumann [5], in these cases the quantum mechanics stops being itself: the operators loose their Hermitianity, transformations loose unitarity. Perhaps, only a multiworld interpretation of quantum transitions by Everett [6] leaves the apparatus of quantum mechanics unperturbed "but at what price!" [7].

At the same time in the theory of quantum transitions a certain case, which is possible to verify experimentally, remains uninvestigated. Let us consider the system "prepared" in the moment $t_1$ with the quantum transition "1" to the future transition "2". Then the system is influenced by an external perturbation and after a period of

----------------

"preparation" $\tau$ the system undergoes quantum transition "2". The moments of both quantum transitions may be registered by the emitted radiation. Having statistical data about the moments of transitions one can calculate the statistical connection between the value of $\tau$ and probability of transition at the moment of time $t_1 + \tau = t_2$. Calculated probability of the transition "2" should contain the moments of the random value $\tau$ as parameters. It is well known from experimental data that the standard methods for calculating transition probabilities are good and there is no necessity to take into account additional parameters connected with $\tau$. Thus, one can make conclusion that probabilities of quantum transitions and the statistical distributions of the $\tau$ value are independent. In other words, for any choice of value of the parameter $\tau$ the distribution of conditional probabilities of quantum transition in moment $t_2$ does not depend on $\tau$. One can ask what $\tau$ values are suitable? The simplest and as one may think, the clear answer is that the $\tau$ values are arbitrary. As it is shown later, such circumstances are emerging if characteristic values of $\tau$ are small or large compared to the time interval characterizing the quantum transition. Such time interval is time of quantum transition between energy levels. We will designate this time interval as T. The value of T depends on the type of quantum transition and is phenomenological in relation to standard quantum mechanics. Author believes that the quantum mechanics apparatus proposed in this work enables to avoid this difficulty.

After proposed by R. Feynman [8] formulation of quantum mechanics by functional integrals it became traditional to consider quantum propagators as a basis for constructing the complex measure in the set of paths of stochastic process, that is identical to Feynman paths [9]. Although the strict mathematical basis for this approach is believed to be a problem far from its solution, the analogies between quantum mechanics apparatus and the theory of stochastic processes may be fruitful.

In this work such analogy is applied. From the theory of diffuse Markov stochastic processes [10] the space of such process $\{x\}$ can be extended to $\{x,u\}$ by introducing new variable in form of the integral of initial process $x(t)$ :

$$u = \int\limits_s^t d\tau A(x(\tau), \tau).$$

Such transformation of stochastic process is one of the methods for calculating the continual functional integrals of stochastic processes. One may construct the direct Kolmogorov equation (or Fokker-Planck equation) to extend process using the method of infinitesimal values [11]. Later we will use this method to broaden the configured space, in which the Schrödinger equation is determined. The properties of solutions of the equation in the broader space enable us to find the parameter that may be interpreted as a time of quantum transition $T$. The final results that may be proved by the experiment are not connected with the method of extended space. They indicate that $T$ value boards the set of $\tau$ in which the quantum transition may be realized. The interpretation proposed in this work suggests that the "quantum windows", which are proportional to $T$ and in which the quantum transition is impossible, will emerge in the space of parameter $\tau$ values

Author hopes what the method of partial wave functions developed in this paper will find various applications in physics, therefore the most of the paper specifically describes the method. The questions related to physical interpretation are placed in the last chapters.

## II. PROPAGATORS "PAST-FUTURE" AND "FUTURE-PAST"

It is known, that the solution $\Psi(x,t)$ of Schrödinger equation for particle:

$$i\eta \frac{\partial \Psi(x,t)}{\partial t} = -\frac{\eta^2}{2m}\Delta\Psi(x,t) + V(x,t)\Psi(x,t) \tag{1}$$

($x$ and $t$ – coordinates of particle and time) being in potential $V(x,t)$, can be written with the aid of propagator $K(x,t|y,s)$, if the wave function $\Psi(x,s)$ is known at the moment $s$:

$$\Psi(x,t) = \int dy K(x,t|y,s)\Psi(y,s) \tag{2}$$

The principle of superposition makes it possible to present solution of the Schrödinger equation in more general form using the set of wave functions $\phi_n(x,s_n)$, that is determined for the arbitrary set of moments of time $s_n$:

$$\Psi(x,t) = \sum_n \int dy K(x,t|y,s_n)\phi_n(y,s_n) \tag{3}$$

In quantum mechanics of reversible processes (it is sufficient to consider the Hamiltonian to be stationary in such cases) every moment of time may be considered as initial. Therefore, propagator $K(x,t|y,s)$ may have Markovian properties (to be more accurate, quasi Markovian properties), because the propagator in description of reversible processes is the complex function determined for $s \leq t$ and for $s \geq t$:

$$K(x,t \mid y,s) = \int dz K(x,t \mid z,r)K(z,r \mid y,s) \tag{4}$$

With the aid of such propagators and the equation (3) one can express the wave function $\psi(x,t)$ by multiple ways that will formally contain contributions of moments $s_n$ that are preceding and post going relatively to $t$.

If the condition of reversibility is not applied, the propagators are considered if $s \le t$; in case of $s > t$, K=0 will be chosen. In this case propagators become the discontinued functions $s$ and $t$. In the present work we will not limit ourselves to $s \le t$ condition for the propagators arguments. The method proposed in this paper should aid one to make conclusions about the contributions of preceding and post going moments of time for non stationary problems as well.

## III.   FIELD FUNCTIONAL INTEGRAL

As it was noticed by R. Feynman, the propagator K may be written in form of continual functional integral with the given complex weight.

$$K(x,t \mid y,s) = \int D[x(\tau)] \exp \left\{ \frac{i}{\hbar} \int_{s}^{t} d\tau \left[ \frac{m}{2} \left( \frac{dx}{d\tau} \right)^2 - V(x(\tau),\tau) \right] \right\} \qquad (5)$$

Feynman's continual integral (3) with element of volume in the space of functions $D[x(\tau)]$ is analyzed in the set of paths that are beginning in the point y at the moment s and coming to the point x at the moment t. The value

$$\exp \left( \frac{im}{2\hbar} \int_{s}^{t} d\tau \left( \frac{dx}{d\tau} \right)^2 \right). \qquad (6)$$

corresponds to the complex weight of path not connected with V(x,t) type of potential energy.

From the structure of continual integral (3) it follows that the parameter $u$, compared to the chosen paths, is considered in the integral construction:

$$u = -\frac{1}{\hbar} \int_{s}^{t} d\tau V(x(\tau),\tau). \qquad (7)$$

In the classical dynamics integral over the potential energy on a path of motion is called generalized virial. In this work we will call the non-dimension value $u$ a functional "field virial" because the value of $u$ is connected with the potential energy of particle in the external field. Considering the summing on paths one can say that the set of paths with fixed $u$ is to be chosen initially, and the summing is done for all values of $u$ afterwards.

In many cases the potential $V[x,t]$ may be treated as having the constant sign. In these cases the sign of the value $u$ may be used for path classification by the initial moment $s$: if $u$ and $V$ have different signs when s<t, the moment of time s is attributed to the past relatively to the moment $t$. If u and V have the same sign when s>t, the moment of time s is attributed to the future relatively to the moment $t$. Formulations in the "paths language" have the higher level of conditionality, so later analyzing the contribution of "past" and "future" on the value of wave function we will analyze the sign of the value $u$.

## IV.   PARTIAL WAVE FUNCTION

Calculating the continual integral over wave function in contradistinction to a propagator it is impossible to determine the initial coordinates and moments of time for paths with definite value of $u$. Nevertheless, knowing the value of $F(u,x,t)$ contribution that corresponds to determined value of $u$ one can express the Schrödinger wave function as follows:

$$\Psi(x,t) = (2\pi)^{-1/2} \int du F(u,x,t) \exp(iu). \tag{8}$$

The function $F$ introduced this way is proposed to be called "partial wave function". (The factor $(2\pi)^{-1/2}$ is introduced in (8) for more convenient representation of the formulas in the further chapters). The differential equation for $F(u,x,t)$ and its solutions are analyzed below.

## V.   DRIFT WAVE EQUATION (DWE)

In the theory of Markov stochastic processes, the standard procedure for constructing differential equations for measuring sets of paths is elaborated. In these equations, named Kolmogorov equations, the quotients at derivatives on coordinates are values of infinitesimal medians and dispersions. In Feynman integrals, instead of non-negative measure used in probability theory, the complex weight of paths is presented. Nevertheless, the analogy with the theory of stochastic processes enables us to indicate the form of equation for $u$ value distribution density $h$:

$$i\hbar \frac{\partial F(u,x,t)}{\partial t} = -\frac{\hbar^2}{2m} \Delta F(u,x,t) + iV(x,t) \frac{\partial F(u,x,t)}{\partial u} \tag{9}$$

We will call this equation a "drift wave equation" (DWE). The right part of the equation can be written with the aid of Hermitian operator $\overset{\wedge}{G}$ :

$$\overset{\backprime}{G} = -\frac{\hbar^2}{2m}\nabla^2 + iV(x,t)\frac{\partial}{\partial u} \tag{10}$$

In analogy with the stochastic processes theory we shall call operator $\overset{\backprime}{G}$ a quantum generating operator. We are also introducing non-dimensional Hermitian operators:

$$\overset{\backprime}{u} = u \text{ and } \overset{\backprime}{\mu} = i\frac{\partial}{\partial u} \tag{11}$$

We will call the first of these operators ($\overset{\backprime}{u}$) a field virial operator and the second, ($\overset{\backprime}{\mu}$) - a scale operator.

## VI.  CONSTRUCTING THE DWE SOLUTIONS

One can propose the general way for solving DWE (9). Firstly, we will examine Fourier-transformation of the partial wave function:

$$\Phi_\mu(x,t) = (2\pi)^{-1/2}\int du\, \exp(i\mu u)F(u,x,t) \tag{12}$$

After using Fourier transformation for the both parts of equation (9) and under condition

$$\exp(i\mu u)F(u,x,t) - \exp(-i\mu u)F(-u,x,t) \to 0 \text{ if } u \to \infty \tag{13}$$

we will obtain equation for Fourier transformation:

$$i\hbar\frac{\partial\Phi_\mu(x,t)}{\partial t} = -\frac{\hbar^2}{2m}\Delta\Phi_\mu(x,t) + \mu V(x,t)\Phi_\mu(x,t) \tag{14}$$

Equation (14) is Schrödinger equation with modified potential $\mu V(x,t)$. If $\mu \neq 0$, this equation can be obtained by substituting $m$ for $\mu m$ and $t$ for $\mu t$ in the initial Schrödinger equation (1). The parameter $\mu$ acts as a scale for potential energy or for mass (it will be called a scale parameter further), therefore the name "scale operator" for $\overset{\backprime}{\mu}$ is justified.

Let the function $\psi_\mu(x,t)$ be a normalized solution of the modified Schrödinger equation (14) with potential $\mu V(x,t)$. Then equation:

$$\Phi_\mu(x,t) = \rho(\mu)\ \psi_\mu(x,t) \tag{15}$$

containing the weight function $\rho(\mu)$ also will be a solution of equation (14). The weight function $\rho(\mu)$ may be chosen with the great extent of arbitrariness restricted only by the following condition: $\rho(\mu) = 0$ out of the space of values of $\mu$, for which the $\psi_\mu(x,t)$ function is definite and continuous. The arbitrariness in the choice of $\rho(\mu)$ corresponds to the arbitrariness in the choice of moments $s_n$ and function $\phi_n(x,s_n)$ in the formula (3).

It remains to find the partial wave function $F(u,x,t)$ that corresponds to particular solution $\rho(\mu)\psi_\mu(x,t)$.

To do that the reverse Fourier transformation can be applied:

$$F(u,x,t) = \left(2\pi\right)^{-1/2} \int d\mu \rho(\mu)\psi_\mu(x,t)\exp(-i\mu u). \tag{15a}$$

The formula (15a) demonstrates that for every solution of Schrödinger equation the whole class of partial wave functions determined by the arbitrariness in choosing the weight function $\rho(\mu)$, may be developed. To construct the partial wave function in a more general form we will examine the set of particular solutions of the equation (9) that can be re-numerated by the factor $k$ (which may be discrete of continuous):

$$f_{\mu,k}(u,x,t) = \left(2\pi\right)^{-1/2} \psi_{\mu,k}(x,t)\exp(-i\mu u). \tag{16}$$

Then, the general solution of DWE appears as a summation by $k$ (in case of discrete parameter $k$) and as an integral over continuous parameter $\mu$:

$$F(u,x,t) = \int d\mu \sum_k \ \rho_k(\mu)\left(2\pi\right)^{-1/2}\psi_{\mu,k}(x,t)\exp(-i\mu u). \tag{17}$$

The transition to the wave function $\psi(x,t)$, according to the formula (8), contains integration over parameter *u*. As a result of integrating *F(u,x,t)* over *u* with factor $\left(2\pi\right)^{-1/2}e^{iu}$ under the sign of integral over $\mu$, $\delta$-function of the form $\delta(\mu-1)$ is obtained:

$$\left(2\pi\right)^{-1/2}\int du F(u,x,t)\exp(iu) = \int d\mu\ \delta(\mu-1)\sum_k\ \rho_k(\mu)\psi_{\mu,k}(x,t). \tag{17a}$$

The result of integrating over $\mu$ is under integral expression, in which $\mu = 1$, that corresponds to the initial value of the mass $m$. As a result the solution of Schrödinger equation (1) for a particle with mass $m$ is obtained.

$$(2\pi)^{-1/2} \int \exp(iu) F(u,x,t) du = \sum_k \rho_k(1) \psi_{1,k}(x,t) \qquad (18)$$

The functions $\psi_{1,k}(x,t)$ by definition coincide with the solutions $\psi_k(x,t)$ of the initial Schrödinger equation.

Let us return to the partial wave function presented in the form (15). One can choose the function $\rho(\mu)$ different from zero in the finite interval $\mu_- < \mu < \mu_+$. In case of restricted function $\varphi(\mu) = \rho(\mu)\psi_\mu(x,t)$ (with fixed $x$ and $t$), according to Riemann lemma [11], we will obtain that

$$F(u,x,t) \to 0 \text{, if} |u| \to \infty . \qquad (19)$$

Nevertheless the condition (19) is sufficient for fulfilling the restriction (13) in initial Fourier transformation (12). This way the finiteness of the interval, in which $\rho(\mu) \neq 0$, is sufficient condition for transition (8) from the partial wave function to Schrödinger equation solution. Of note, in this case:

$$F(u,x,t) \to 0 \text{, if} |t| \to \infty \qquad (19a)$$

## VII.  DWE AND SCHRÖDINGER EQUATION

Let us examine the necessary and sufficient condition, under which the transformation (8) applied to the solution of the DWE (9) really leads to the solution of Schrödinger equation. Using the operation (8) for both parts of the equation (9), one obtains an equation in the space $(x,t)$, which is identical to the Schrödinger equation (1) if in the space where $V(x,t) \neq 0$, the following condition is fulfilled:

$$\int du \exp(iu) F(u,x,t)) = i \int du \exp(iu) \frac{\partial}{\partial u} F(u,x,t). \qquad (20)$$

This condition one may present as follows:

$$\int du [F(u,x,t) - i \frac{\partial}{\partial u} F(u,x,t)] \exp(iu) = 0 \qquad (21)$$

If to require each under integral expression to equal zero for each value of $u$ one will have to consider functions $F(u,x,t) = \psi(x,t)\exp(-iu)$. This is particular solution of equation (9) for which operation (8) gives the diverging result. Nevertheless if one uses the general solution in form of an integral (15) the left part of the equation (21) takes form:

$$(2\pi)^{-1/2} \int du \int d\mu (1-\mu)\rho(\mu)\psi_\mu(x,t)\exp(iu - i\mu u). \qquad (22)$$

The under integral expression in (22) does not equal zero but the integral over $u$ equals zero:

$$(2\pi)^{1/2} \int d\mu \ (1-\mu) \ \rho(\mu) \ \delta(1-\mu) \ \psi_\mu(\mu,x,t) = 0. \qquad (23)$$

This means that solutions in the form of an integral (15) fulfills the necessary and sufficient condition (21) in case of the broad assumptions about the form of the weight function.

# VIII. THE SIMPLEST PROPERTIES OF DWE SOLUTIONS

8.1. If $F(u,x,t)$ is the solution of DWE (9), the function $F_1(u,x,t) = F(u,x,t) + K$ is also the solution ($K$ is an arbitrary constant). Therefore, if $F(u,x,t)$ is transformed into a solution of Schrödinger equation by formula (8), then for $K \neq 0$ the result of transformation contains an additional contribution $V.p.K \int du \exp(iu)$, which oscillates as a function of the upper limit of integration. Summing up the condition of transformability enables us to choose no more than one solution from the set of functions of the $F(u,x,t) + K$ form.

8.2. The value of the potential $V(x,t)$ does not depend upon a field phase $u$ (otherwise the operator $\overset{\downarrow}{G}$ becomes non-Hermitian). Therefore, if $F(u,x,t)$ is the solution of DWE, the function $F_1(u,x,t) = F(u+C,x,t)$ is the solution as well. Here $C$ is an arbitrary constant. If the transformation (8) gives for $F(u,x,t)$ the result $R(F(u,x,t)) = \Psi(x,t)$, then for $R(F(u,x,t)) = \Psi(x,t)$ $R(F_1(u,x,t)) = \Psi(x,t)\exp(-iC)$ is correct.

8.3. Instead of the constant value of the shift $C$ one may consider the arbitrary function of time $\hbar^{-1}\varphi(t)$. If $F(u,x,t)$ is the solution of the equation (9), the function $F_1(u,x,t) = F(u+\hbar^{-1}\varphi(t),x,t)$ is the solution for the new DWE, in which the potential $V(x,t)$ is substituted by the $V(x,t) + d\varphi(t)/dt$. Such substitution corresponds to the shift in counting off the values of the potential.

# IX. THE BALANCE EQUATION FOR PARTIAL WAVE FUNCTIONS

We can analyze properties of DWE solutions with the nonstationary potential using the integral properties of the partial wave function. Let us examine two DWE solutions $F_1(u,x,t)$ and $F_2(u,x,t)$, differing for instance in choice of their weight functions $\rho_1(\mu)$ and $\rho_2(\mu)$ in (15). The pair of DWE solutions in (9) satisfies the following integral equation, which can be called a "balance equation":

$$\frac{\partial}{\partial t}\int dx F_1^*(u,x,t)F_2(u,x,t) = \frac{\partial}{\hbar \partial u}\int dx F_1^*(u,x,t)V(x,t)F_2(u,x,t)$$

(24)

To obtain the balance equation (24) let us write DWE (9) for $F_2$ and complex-conjugate equation for $F_1^*$. Let us multiply the both parts of the first equation by $F_1$ and both parts of the second by $F_2$:

$$i\hbar F_1^*(u,x,t)\cdot\frac{\partial F_2(u,x,t)}{\partial t} = F_1^*(u,x,t)\cdot\hat{T}F_2(u,x,t) +$$

$$+iF_1^*(u,x,t)\hat{V}(x,t)\frac{\partial F_2(u,x,t)}{\partial u},$$

(25)

$$-i\hbar F_2(u,x,t)\frac{\partial F_1^*(u,x,t)}{\partial t} = F_2(u,x,t)(\hat{T}F_1(u,x,t))^* -$$

$$-iF_2(u,x,t)\hat{V}^*(x,t)\frac{\partial F_1^*(u,x,t)}{\partial u}.$$

(26)

Here the $\hat{T}$ - Hermitian operator for kinetic energy. Let us subtract one of the obtained equations from another one and integrate both parts on space variables $x$:

$$i\hbar\frac{\partial}{\partial t}\int dx F_1^*(u,x,t)F_2(u,x,t) =$$

$$i[\int dx(\hat{V}^*(x,t)F_1^*(u,x,t))\frac{\partial}{\partial u}F_2(u,x,t)dx +$$

$$\int dx F_2(u,x,t)\frac{\partial}{\partial u}\hat{V}^*(x,t)F_1^*(u,x,t)dx] +$$

$$\int dx\left(F^*(u,x,t)\hat{T}F(u,x,t) - F(u,x,t)(\hat{T}F(u,x,t))^*\right).$$

(27)

The last integral in the right part of the equation (27) equals zero because $\hat{T}$ is the Hermitian operator. Because the operator $\hat{V}$ is Hermitian we obtain:

$$i\hbar\frac{\partial}{\partial t}\int dx F_1^*(u,x,t)F_2(u,x,t)=i\frac{\partial}{\partial u}\int dx F_1^*(u,x,t)\hat{V}(x,t)F_2(u,x,t).$$  (27a)

Thus, the balance equation (24) is obtained.

# X. NORMALIZATION IN THE SPACE $\{u,x\}$ AND $\{t,x\}$

10.1. From DWE one can obtain several conclusions which enable us to consider parameters $u$ and $t$ as mutually substitutive "times". One may use these parameters for parameterization of the evolution of the partial wave function in the space $\{u,x\}$. Indeed, the conservation of the norm of partial wave function in case of change of time $t$ follows from the Hermitianity of the operator $\hat{G}$ (10). For this purpose is sufficient to consider consequences of the equation (24) if $F_1(u,x,t)=F_2(u,x,t)=F(u,x,t)$:

$$\frac{\partial}{\partial t}\int du\int dx F^*(u,x,t)F(u,x,t)=0.$$  (28)

It is possible to introduce dimensionless invariant $N$, which not time-dependent

$$N=\int\limits_{-\infty}^{\infty} du\int dx\left|F(u,x,t)\right|^2.$$  (29)

10.2. Examining the Hermitian operator $\hat{D}=\frac{\hbar^2}{2m}\Delta+i\hbar\frac{\partial}{\partial t}$ and re-writing with its aid the equation (9) in the form: $iV(x,t)\frac{\partial F(u,x,t)}{\partial u}=\hat{D}F(u,x,t)$ we obtain the conservation of normalizing in the space $\{t,x\}$:

$$\frac{\partial}{\partial u}\int\limits_{-\infty}^{\infty} dt\int dx V(x,t)\left|F(u,x,t)\right|^2=0$$  (30)

It follows from the equation (29) that the value $u$ has the characteristics that differentiate the time variance from the space coordinates. As well as for the time $t$, the analog of the conservation of wave function normalization can be introduced for $u$. If one changes $dx$ for $V(x,t)dx$ in the space $\{t,x\}$ it is possible to introduce the non-dimensional integral invariant, which is independent from the field phase $u$:

.

$$M = \frac{1}{\mathrm{h}} \int\limits_{-\infty}^{\infty} dt \int dx V(x,t) \left| F(u,x,t) \right|^2 \tag{31}$$

Introduction of an additional weight $V(x,t)$ to the under integral expression corresponds to substitution in one-dimensional problem of classical mechanics but in the quantum case the one-dimensionality is not necessary (see the Supplement 1).

**10.3.** The existence of invariant (31) enables us to draw conclusion about classification of paths in a non-stationary case. Let us consider a case when potential $V(x,t)$ has constant sign. From equation (30) in it follows then that the invariant $M$ is different from zero if the values of $u$ have different signs. According to the classification used in this article, this means that in the course of evolution of the quantum system the paths beginning in the forestall moments have made contributions to the particle wave function at least sometimes.

**10.4.** The Hermitianity of operators $\overset{\downarrow}{G}$ и $\overset{\downarrow}{D}$ means fulfillment of the conditions of sufficiently fast decrease of $F(u,x,t)$ if $u \to \infty$ or $t \to \infty$. Let us suppose that in the equations (28) and (30) the integrals over $u$ and $t$ are considered the main values. In this case the sufficient conditions, under which the equations (28) and (30) are fulfilled, correspondingly are:

$$\int dx V(x,t) \left| F(u,x,t) \right|^2 - \int dx V(x,t) \left| F(-u,x,t) \right|^2 \to 0 \text{, if } u \to \infty \text{,} \tag{32}$$

$$\int dx \left| F(u,x,t) \right|^2 - \int dx \left| F(u,x,-t) \right|^2 \to 0 \text{, if } t \to \infty \text{.} \tag{32a}$$

Direct calculations in Chapter 13 and Supplement 3 show that equations (28) and (30) are correct for the case of stationary potential; the values of invariants $N$ and $M$ are calculated for this case as well.

**10.5.** The consequence of equation (27a) (if the conditions analogous to (32) are justified) is an existence of two real (but not obligatory with constant sign) values that do not depend on time:

$$S = 1/2[\int du dx F_1^*(u,x,t) F_2(u,x,t) + \int du dx F_1(u,x,t) F_2^*(u,x,t)], \qquad (33)$$

$$A = i/2[\int du dx F_1(u,x,t) F_2^*(u,x,t) - \int du dx F_1^*(u,x,t) F_2(u,x,t)]. \qquad (34)$$

If $F_1(u,x,t) = F_2(u,x,t) = F(u,x,t)$, the invariants $N$ and S mutually coincide.

# XI. PARTIAL WAVE FUNCTIONS AND PHYSICAL VALUES

The wave function $\Psi(x,t)$ represents the intermediate «informational» mathematical object in quantum mechanics. To establish correspondence between the wave function and the results of the measurements, the Hermitian operators of the physical values $\hat{A}$ with the full sets of orthonormalized eigenfunctions $a_n(x)$ with the eigenvalues $A_n$ (if the spectrum of operator $\hat{A}$ is discrete) being introduced. The projection $\Psi(x,t)$ on $a_n(x)$ defines the probability of detection $P_n$ of the value $A_n$ by the formula $P_n(t) = \left| \langle a_n | \psi \rangle \right|^2$. These calculations require transition from the partial functions to the regular wave function $\Psi(x,t)$. But the task of calculating the values $P_n$ can be formulated using the partial functions of special form. Let us consider that the functions $a_{\mu,n}(x)$ represent the full set of the orthonormalized eigenfunctions of the operator $\hat{A}_\mu$ with the arbitrary value of the parameter $\mu$. One can examine the partial wave function being written using the normalized solution $\psi_\mu(x,t)$ of the equation (14).

$$F(u,x,t) = (2\pi)^{-1/2} \int d\mu \rho(\mu) \psi_\mu(x,t) \exp(-i\mu u). \qquad (35)$$

$F$ can be represented in the following form, using the functions $a_{\mu,n}(x)$:

$$F(u,x,t) = (2\pi)^{-1/2} \sum_n \int d\mu \rho(\mu) a_{\mu,n}(x)$$
$$< a_{\mu,n}(x)|\psi_\mu(x,t) > \exp(-i\mu u).$$

(36)

The possibility to introduce the probability distribution with normalization which does not depend on time is based on existence of bilinear invariants of $N$ type (29). Let us look at the case when the partial wave function (35) is normalized by 1. It is sufficient to normalize the functions $\psi_\mu(x,t)$ и $\rho(\mu)$ the same way. By computing N we obtain:

$$\sum_n \int d\mu |\rho(\mu)|^2 \left|< a_{\mu,n}(x)|\psi_\mu(x,t) >\right|^2 = 1.$$

(37)

Each term of this sum can be interpreted as the determination of the probability $\overset{\circ}{P}_n(t)$ to detect the value $A_n$:

$$\overset{\circ}{P}_n(t) = \int d\mu |\rho(\mu)|^2 \left|< a_{\mu,n}(x)|\psi_\mu(x,t) >\right|^2.$$

(38)

The determinations of $P_n$ and $\overset{\circ}{P}_n$ do not agree in general, although they are connected by the limiting transition. By choosing:

$$|\rho(\mu)|^2 = \delta(\mu - 1)$$

(39)

we will have $P_n = \overset{\circ}{P}_n$. As a result of all calculations with the partial wave functions we have to leave only those that do not depend on the choice of the weight functions.

# XII. UNCERTAINTY "$\mu$–$U$"

Thanks to conservation of the norm in the space $\{u, x\}$, the value $\left| F(u, x, t) \right|^2$ can be considered as the distribution function for $\{u, x\}$. The calculation of the mean values for such distributions, normalized by one, amounts to the integration over $x$ and $u$. We will use double angle brakes to introduce mean for calculation of the mean values $\left\langle \left\langle u^k \right\rangle \right\rangle$ and $\left\langle \left\langle \mu^k \right\rangle \right\rangle$ with fixed $t$. The dispersion $D_\mu$ for the value $\mu$ is found by the standard formula:

$$D_\mu = \left\langle \left\langle \mu^2 \right\rangle \right\rangle - \left\langle \left\langle \mu \right\rangle \right\rangle^2 . \tag{40}$$

Similarly we define the dispersion $D_u$ for the parameter u. The operators $\hat{\mu}$ and $\hat{u} = u$ are noncommuting:

$$\hat{\mu}\hat{u} - \hat{u}\hat{\mu} = i . \tag{41}$$

According to the general theory about noncommuting Hermitian operators, one obtains the "μ–u" uncertainty relation:

$$(\left\langle \left\langle \mu^2 \right\rangle \right\rangle - \left\langle \left\langle \mu \right\rangle \right\rangle^2)(\left\langle \left\langle u^2 \right\rangle \right\rangle - \left\langle \left\langle u \right\rangle \right\rangle^2) \geq \frac{1}{4} . \tag{42}$$

# XIII. STATIONARY POTENTIAL. THE PARTIAL WAVE EIGENFUNCTIONS

Let us consider the DWE (9) for the case when the interaction's potential does not depend on time explicitly. The eigenfunctions $g_{\mu,n}(u, x)$ of the operator $\hat{G}$, which corresponds to the eigenvalues of $G_{\mu,n}$, have a form:

$$g_{\mu,n}(u, x) = \left( 2\pi \right)^{-1/2} \psi_{\mu,n}(x) \exp(-i\mu u). \tag{43}$$

Functions $\psi_{\mu,n}(x)$ correspond to the following stationary Schrödinger equation:

$$-\frac{\hbar^2}{2m}\nabla^2\psi_{\mu,n}(x) + \mu V(x)\psi_{\mu,n}(x) = G_{\mu,n}\psi_{\mu,n}(x) \quad . \qquad (44)$$

The equation (44) is convenient to present in the form of the stationary Schrödinger equation for the particle with the mass $\mu\,m$ :

$$-\frac{\hbar^2}{2\mu m}\nabla^2\psi_{\mu,n}(x) + V(x)\psi_{\mu,n}(x) = E_{\mu,n}\psi_{\mu,n}(x) \quad . \qquad (45)$$

In the equation (45) $E_{\mu,n}$ is the n-th energy level for the particle with the mass $\mu\,m$ . Comparing the equations (44) and the (45) one finds:

$$G_{\mu,n} = \mu E_{\mu,n} \quad . \qquad (46)$$

Further we will examine the localized solutions of the equation (44) and consider the functions $\psi_{\mu,n}(x)$ orthonormalized if the $\mu$ is fixed. In this case, the functions $g_{\mu,n}(u,x)$ are normalized by the following condition:

$$\int du \int dx g^*_{\mu_1,n_1}(u,x) g_{\mu_2,n_2}(u,x) = \delta(\mu_1 - \mu_2)\delta_{n_1,n_2} \quad . \qquad (47)$$

## XIV. TRANSITION TO THE WAVE FUNCTIONS, $\mu$–PACKETS

The solution of the drift wave equation can be constructed as a linear combination of the eigenfunctions $g_{\mu,n}$ . The integration over the continuous quantum number $\mu$ and the summation over $n$ are carried out, if the solution of the stationary Schrödinger equation leads to the quantum number spectrum:

$$F(u,x,t) = \sum_n a_n \int d\mu \rho_n(\mu) g_{\mu,n}(u,x) \exp\{-i\hbar^{-1}G_{\mu,n}t\}. \tag{48}$$

To make the expression (48) for $F(u,x,t)$ meaningful, let us consider $\rho_n(\mu) = 0$ outside the limits of the parameter $\mu$, for which the function $f_n(\mu) = E_{\mu,n}$ is defined and continuous. Choosing the functions $\rho_n(\mu)$ as not equal to zero within the limited intervals $\mu_{n-} < \mu < \mu_{n+}$, for the functions $\psi_{\mu,n}(x)$ limited by the set, according to Riemann lemma we will have $F(u,x,t) \to 0$ as $|u| \to \infty$. Therefore, the constricted solutions obey the sufficient condition (10) of the transition to Schrödinger equation. If we integrate F(u,x,t) (48) over $u$ with the factor $(2\pi)^{-1/2} \exp(iu)$, then the factor $\delta(\mu-1)$ appears under the integral sign over $\mu$. Performing the integration over $\mu$ we obtain the expression, which corresponds to the concrete value of the mass $m$:

$$\Psi(x,t) = (2\pi)^{-1/2} \int \exp(iu)F(u,x,t)du =$$
$$= \sum_n a_n \rho_n(1)\psi_{1,n}(x)\exp\{-i\hbar^{-1}E_{1,n}t\}. \tag{49}$$

By definition, the functions $\psi_{1,n}(x)$ and the energies $E_{1,n}$ coincide with the eigenfunctions $\psi_n(x)$ and the levels of the energy for $E_n$ initial Schrödinger equation.

The solutions of the stationary DWE (9) of the form that follows we will call «$\mu$–packets»:

$$F_n(u,x,t) = \frac{1}{\sqrt{2\pi}} \int d\mu \rho_n(\mu) \psi_{\mu,n}(x)\exp\{-i\mu(\hbar^{-1}E_{\mu,n}t+u)\}. \tag{50}$$

## XV. ORTHONORMALITY OF THE PARTIAL WAVE FUNCTIONS

Let us consider the orthonormality of the solutions of the DWE. Thanks to the condition of the orthonormality (47) for the eigenfunctions $g_{\mu,n}$ the μ–packets of the different indexes n, *k* are orthogonal:

$$\int du dx F_n^*(u,x,t) \cdot F_k(u,x,t) = 0 \qquad (51)$$

According to (40) the inner product of the $\mu$–packets with the same indexes $n$ is equal to:

$$\int du dx F_n^*(u,x,t) F_n(u,x,t) = \int d\mu \left| \rho_n(\mu) \right|^2 \qquad (52)$$

The equations (45) and (46) mean that for the linear combinations of the μ-packets the norm in the space $\{u,x\}$ is invariant. The invariant $N$ (28a) for the partial wave function (50) has the form:

$$N = \sum_n \left| a_n \right|^2 \int d\mu \left| \rho_n(\mu) \right|^2 \qquad (53)$$

As usual, we will normalize the μ-packets by one. For this it is necessary and sufficient that:

$$\int d\mu \left| \rho_n(\mu) \right|^2 = 1 \qquad (54)$$

It is convenient to represent $\rho_n(\mu)$ in terms of the real functions $r_n(\mu)$ and $\varphi_n(\mu)$:

$$\rho_n(\mu) = r_n(\mu) \exp\left[ i\varphi_n(\mu) \right] \qquad (55)$$

Then the invariant $S$ (31) has the form:

$$S = \int d\mu r_1(\mu) r_2(\mu) \cos(\varphi_1(\mu) - \varphi_2(\mu)) \qquad (56)$$

for each two μ–packets with the different $r_{1,2}(\mu)$ and $\varphi_{1,2}(\mu)$

## XVI. SPACES $\{u, x\}$. UNCERTAINTY «$g - t$»

We can calculate the mean values $t^k$ for fixed $u$ if we use the normalized function $F(u, x, t)$:

$$\left\langle \left\{ t^k \right\} \right\rangle = \text{sgn}(V) \int dx \int\limits_{-\infty}^{\infty} dt V(x) |F(u, x, t)|^2 t^k \ . \tag{57}$$

In this case the double brackets $\langle \{...\} \rangle$ indicate the mean value on $x$ and on $t$. The results of such calculations satisfy the principle of uncertainty, which is analogous to Eq. (42). Let us consider the parameter $g$, which takes the values at the set of eigenvalues $G_{\mu,n}$ of the generating operator $\hat{G}$. The value $g$ is included in the combination $\exp\left(-\dfrac{i}{\text{h}} gt\right)$ in the expression of the function $F_n(u, x, t)$. The last function is the eigenfunction of the Hermitian operator $\overset{)}{g}$:

$$\overset{)}{g} = i\text{h} \frac{\partial}{\partial t} \ , \quad \overset{)}{g}t - t\overset{)}{g} = i\text{h}. \tag{58}$$

Let's note that the introducing of the additional weight $V(x)$ in the definition of the innerproduct of functions F(u,x,t) conserve the Hermitianity of the operator, which is acting only on the variable $t$. The relation

$$\left[ \left\langle \left\{ t^2 \right\} \right\rangle - \left( \left\langle \{t\} \right\rangle \right)^2 \right] \left[ \left\langle \left\{ g^2 \right\} \right\rangle - \left( \left\langle \{g\} \right\rangle \right)^2 \right] \geq \frac{\text{h}^2}{4} \tag{59}$$

is following from the Eq. (84).

Here the mean value $g^k$ is defined for the linear combinations of the μ-packets with the constant sign value $\text{sgn}(V)$ :

$$\left\langle \left\{ g^k \right\} \right\rangle = \text{sgn}(V) \int dx \int dt V(x) F^*(u,x,t) \left( \overset{\circ}{g} \right)^k F(u,x,t) =$$

$$= \sum_n |b_n|^2 \int\limits_{\mu-}^{\mu+} d\mu \left| \rho_n(\mu) \right|^2 \left( \mu E_{\mu,n} \right)^k .$$

(60)

# XVII. NONSTATIONARY DRIFT WAVE EQUATION (NDWE)

17.1. The general properties of the solutions of the nonstationary drift wave equations were studied in sections **8-10**. The analysis of the such solutions by the method of the perturbation theory of will be represented below.

First, let us consider the conditions we should impose on the solution $F(u,x,t)$ of the nonstationary DWE to be able to present it by means of the solutions $F_n(u,x,t)$ of some stationary DWE. to represent it using the set of the solutions $F_n(u,x,t)$ of any stationary drift wave equation. The function $F(u,x,t)$ of rather general type can be represented using the transformation:

$$F(u,x,t) = (2\pi)^{-1/2} \int d\mu \Upsilon_\mu(x,t) \exp(-i\mu u),$$

(61)

$$\Upsilon_\mu(x,t) = (2\pi)^{-1/2} \int du \exp(i\mu u) F(u,x,t).$$

(62)

In these general formulas the range of the parameter $\mu$ is arbitrary. Now we restrict the set of functions $\Upsilon_\mu(x,t)$ by the condition that the range of parameter $\mu$, where $\Upsilon_\mu(x,t) \neq 0$, is the subset of the values of parameter $\mu$, for which the orthonormalized functions $\psi_{\mu,n}(x)$ of some stationary problem with the potential $V(x)$ are defined. The unique expression:

$$\Upsilon_\mu(x,t) = \sum_n C_n(\mu,t) \psi_{\mu,n}(x)$$

(63)

is valid under the condition above and for the fixed *t*.

The coefficients $C_n(\mu, t)$ are given by:

$$C_n(\mu, t) = \int dx \psi^*_{\mu,n}(x) \Upsilon_\mu(x, t). \tag{64}$$

are it's coefficients.

Now the initial function can be represents as the sum:

$$F(u, x, t) = (2\pi)^{-1/2} \sum_n \int d\mu \exp(-i\mu u) C_n(\mu, t) \psi_{\mu,n}(x). \tag{65}$$

We can rewrite the latter equation by using the notations from section **13** for the $\mu$-packets.

In order that we have to represent the functions $C_n(\mu, t)$ as a product

$$\begin{aligned} C_n(\mu, t) &= b_n(\mu, t) \rho_n(\mu) \exp(-i\mu \mathrm{h}^{-1} E_{\mu,n} t) = \\ &f_n(\mu, t) \exp(-i\mu \mathrm{h}^{-1} E_{\mu,n} t) \end{aligned}. \tag{66}$$

Here $\rho_n(\mu)$ is the normalized weight function chosen above, the value $E_{\mu,n}$ is defined by the choice of the potential *V(x)*. The potential *V(x)* is usually chosen according to the condition that the functions $b_n(\mu, t)$ is changing by the time slowly compared to $\exp(-i\mu \mathrm{h}^{-1} E_{\mu,n} t)$. The choice of the functions $\rho_n(\mu)$ is subjected to condition: $\rho_n(\mu) \neq 0$ for those values of $\mu$, for which $\Upsilon_\mu(x, t) \neq 0$ with arbitrary values of *t*.

**17.2.** It is convenient to calculate the mean value of parameter *u* for the arbitrary presentation of the function $F(u, x, t)$ in the row (64):

$$\langle u \rangle = \sum_n A_n + \sum_{m \neq n} B_{mn}. \tag{67}$$

Here

$$A_n = \text{Im} \int d\mu (f_{\mu,n}(t) e_{\mu,n}(t))^* \frac{\partial}{\partial \mu} f_{\mu,n}(t) e_{\mu,n}(t) +$$

$$+ \text{Im} \int d\mu \int dx \left| f_{\mu,n}(t) \right|^2 \psi^*_{\mu,n}(x) \frac{\partial}{\partial \mu} \psi_{\mu,n}(x), \qquad (68)$$

$$B_{m,n} = 2 \text{Im} \int d\mu (f_{\mu,n}(t) e_{\mu,n}(t))^* f_{\mu,m}(t)$$

$$e_{\mu,m}(t) \left\langle \mu, n \left| V \right| \mu, m \right\rangle \left[ \mu (E_{\mu,m} - E_{\mu,n}) \right]^{-1}$$

$$= 2 \text{Im} \int d\mu (f_{\mu,m}(t) e_{\mu,nm}(t))^* f_{\mu,n}(t)$$

$$e_{\mu,n}(t) \left\langle \mu, m \left| V \right| \mu, n \right\rangle \left[ \mu (E_{\mu,n} - E_{\mu,m}) \right]^{-1}, \qquad (69)$$

$$e_{\mu,n}(t) = \exp(-i\mu \hbar^{-1} E_{\mu,n} t). \qquad (70)$$

**17.3.** The theory of perturbation, which can be used for the partial wave functions, is analogous to the one being developed for the wave functions. Let us examine a non-stationary DWE, in which a non-stationary part $W(x,t)$ is added to a stationary potential $V(x)$ from the moment $t_1$. In the following, this non-stationary part will be considered a small contribution. If $t_1 < t$ DWE takes form:

$$i\hbar \frac{\partial F(u,x,t)}{\partial t} = -\frac{\hbar^2}{2m} \Delta F(u,x,t) + i(V(x,t) + W(x,t)) \frac{\partial F(u,x,t)}{\partial u}. \qquad (71)$$

Let us select as an initial state ($t < t_1$) a normalized $\mu$-packet

$$F_s(u,x,t) = \frac{1}{\sqrt{2\pi}} \int d\mu r_s(\mu) \psi_{\mu,s}(x) \exp\{i\varphi_s(\mu) - i\mu(\text{h}^{-1}E_{\mu,s}t + u)\}. \tag{72}$$

Then $t_I < t$ according to a theory of perturbations for DWE the following solution will be obtained:

$$F(u,x,t) = (2\pi)^{-1/2} \sum_n \int d\mu\, b_{n,s}(\mu,t) r_s(\mu) \psi_{\mu,n}(x)$$
$$\exp(i\varphi_s(\mu) - i\mu \text{h}^{-1} E_{\mu,n} t - i\mu u). \tag{73}$$

$$b_{s,s}(\mu,t) = 1 - \frac{i\mu}{\text{h}} \int_{t_I}^{t} d\tau <\mu,s|W(\tau)|\mu,s> -$$
$$\frac{\mu^2}{\text{h}^2} \int_{t_I}^{t} d\tau_2 <\mu,s|W(\tau_2)|\mu,s> \int_{t_I}^{\tau_2} d\tau_1 <\mu,s|W(\tau_1)|\mu,s> -$$
$$\frac{i\mu^2}{\text{h}^2} \sum_{n\neq s} \int_{t_I}^{t} d\tau_2 <\mu,s|W(\tau_2)|\mu,n> \exp(\frac{i\mu\tau_2}{\text{h}}(E_{\mu,s} - E_{\mu,n}))$$
$$\int_{t_I}^{\tau_2} d\tau_1 <\mu,s|W(\tau_1)|\mu,s> \exp(\frac{i\mu\tau_1}{\text{h}}(E_{\mu,n} - E_{\mu,s})), \tag{74}$$

$$b_{n\neq s,s}(\mu,t) = -\frac{i\mu}{\text{h}} \int_{t_I}^{t} d\tau <\mu,n|W(\tau)|\mu,s> \exp(\frac{i\mu\tau}{\text{h}}(E_{\mu,n} - E_{\mu,s})). \tag{75}$$

Coefficient $b_{s,s}(\mu,t)$ is calculated in the second order of the perturbation theory.

# XVIII.  HELLMAN-FEYNMAN  FORMULAS  FOR  MASS VARIABILITY

**18.1.** The following theorem, which proof has been shown in the Supplement 2, will be useful in further calculations.

Let eigenfunctions $\psi_{\lambda,l}(x)$ and $\psi_{\nu,n}(x)$ of the stationary Schrödinger's equation (38) that have an effective mass $\mu m$ correspond to the two values of $\mu$ ($\mu_1 = \lambda$ и $\mu_2 = \nu$) and two energy levels with the numbers $l$ and $n$. Then the following equation is correct:

$$(\lambda - \nu)\langle \lambda, l \mid V \mid \nu, n \rangle = (\lambda E_{\lambda,l} - \nu E_{\nu,n})\langle \lambda, l \mid \nu, n \rangle, \qquad (76)$$

with the following designations used

$$\langle \lambda, l \mid \nu, n \rangle = \int dx \, \psi_{\lambda,l}^{*}(x) \psi_{\nu,n}(x), \qquad (77)$$

$$\langle \lambda, l \mid V \mid \nu, n \rangle = \int dx \, \psi_{\lambda,l}^{*}(x) V(x) \psi_{\nu,n}(x). \qquad (78)$$

**18.2.** From the equation (76) two consequences, useful for further calculations, can be deduced. First, for the coinciding eigenvalues of the operator $\overset{\lambda}{G}$, i.e. if

$$\lambda E_{\lambda,l} = \nu E_{\nu,n}, \qquad (79)$$

but with different parameters $\lambda$ and $\nu$, the following equality should be correct:

$$\langle \lambda, l \mid V \mid \nu, n \rangle = 0. \qquad (80)$$

Second, if $l = n$, one can analyze the limiting transition $\lambda - \nu = \Delta \nu \to 0$. Having divided both parts of the equality (76) by $\Delta \nu$, one will obtain at the limit:

$$\langle \nu, n \mid V \mid \nu, n \rangle = \langle \nu, n \mid \nu, n \rangle \frac{\partial}{\partial \nu}\left(\nu E_{\nu,n}\right). \qquad (81)$$

Introducing designation :

$$V_{\mu,n} = \langle \mu, n \mid V \mid \mu, n \rangle, \qquad (82)$$

and considering that for the normalized functions $\langle \mu, n \mid \mu, n \rangle = 1$, the equality (81) can be re-written in the following form:

$$V_{\mu,n} = \frac{\partial}{\partial \mu}(\mu E_{\mu,n}) \ . \tag{83}$$

This equality can be viewed as a particular example of the known theorem about eigenvalues of a Hamiltonian, which has a scalar parameter.

**18.3.** Stationary Schrödinger's equation for $\psi_{\mu,n}$ can be written as:

$$\left( \overset{l}{T} + \mu V \right) \psi_{\mu,n} = \mu E_{\mu,n} \psi_{\mu,n} . \tag{83a}$$

Using designations (82), it follows from the last equality:

$$\langle \mu, n \mid \overset{l}{T} \mid \mu, n \rangle = \mu E_{\mu,n} - \mu V_{\mu,n} \tag{84}$$

Hence, considering (83), one obtains:

$$\langle \mu, n \mid \overset{\rightarrow}{T} \mid \mu, n \rangle = -\mu^2 \frac{\partial}{\partial \mu} E_{\mu,n} \tag{85}$$

Let us consider that for the localized states $\langle \mu, n \mid \overset{l}{T} \mid \mu, n \rangle > 0$, from which it follows that for the localized states

$$\frac{\partial}{\partial \mu} E_{\mu,n} < 0 \ . \tag{86}$$

**18.4.** Designation (83) is convenient to use analyzing normalization in the space $\{u, x\}$ in case of stationary potential. For the localized states in the stationary potential of an integral the preservation of the integral (30) can be tested by direct calculations (see Supplement 3). It is assumed in these calculations that $V_{\nu,n}$ and $V_{\lambda,l}$ maintain the sign, which we designate as $\mathrm{sgn}(V)$. In this case for the levels $n$ and $l$ integral:

$$M_{n,l}(u) = \frac{1}{\eta} \int dx \int dt V(x) F_n^*(u,x,t) F_l(u,x,t) \tag{87}$$

becomes equal to:

$$M_{n,l} = \mathrm{sgn}(V) \delta_{n,l} \ . \tag{88}$$

Of note, the last result does not depend on the weight function choice.

# XIX.  MOMENTS OF THE FIELD VIRIAL

**19.1.** The first moments of the field virial $u$ belong to the uncertainties correlation (35). Let us calculate them for a normalized $\mu$- packet. Let us introduce designation $\left\langle\left\langle n\mid u^k\mid n\right\rangle\right\rangle$ for an average value $u^k$ by the $\mu$-packet $F_n(u,x,t)$:

$$\left\langle\left\langle n\mid u^k\mid n\right\rangle\right\rangle = \int du \int dx F_n^*(u,x,t)u^k F_n(u,x,t),, \tag{89}$$

Calculating an integral by $u$ under an integral by $\mu$ and $\nu$ signs, the factor appears:

$$. \qquad \frac{i}{2}(\frac{\partial}{\partial\mu}-\frac{\partial}{\partial\nu})\delta(\mu-\nu) \tag{89a}$$

In order to omit details connected with the marginal contributions while integrating by the interval $[\mu_-,\mu_+]$, the distribution $\rho(\mu)$, subjected to the following condition

$$\rho(\mu_+)=\rho(\mu_-)=0: \tag{90}$$

will be used in further calculations.

For the diagonal matrix element of the potential, calculated for the wave function $\psi_{\mu,n}(x)$, it is convenient to use designation (55). Eigenfunctions $\psi_{\mu,n}$ can be selected as real. It is also convenient to write $\rho_n(\mu)$ via the real functions $r_n(\mu)$ and $\varphi_n(\mu)$

$$\rho_n(\mu)=r_n(\mu)\exp\left[i\varphi_n(\mu)\right]. \tag{91}$$

Considering (90), one will obtain:

$$\left\langle\left\langle n\mid u\mid n\right\rangle\right\rangle = \int d\mu r_n^2 \frac{\partial}{\partial\mu}\varphi_n - \frac{t}{h}\int d\mu r_n^2 V_{\mu,n} \quad . \tag{92}$$

In this case the phase of the weight function $\rho_n(\mu)$ defines the moment of time $t_{0n}$, for which the condition $\left\langle\left\langle n\mid u\mid n\right\rangle\right\rangle=0$ is fulfilled.

$$t_{0n}=h\int d\mu r_n^2 \frac{\partial}{\partial\mu}\varphi_n / \int d\mu r_n^2 V_{\mu,n}, \tag{93}$$

$$\left\langle\left\langle n\,|\,u\,|\,n\right\rangle\right\rangle = -\frac{1}{h}\left(t-t_{0n}\right)\int d\mu\, r_n^2 V_{\mu,n}. \tag{94}$$

In the following the $\mu$ - packets with

$$\varphi_n(\mu) = \mu h^{-1} t_{0n}\int d\mu\, r_n^2 V_{\mu,n}. \tag{95}$$

will be used.

**19.2.** Calculating $\left\langle\left\langle n\,|\,u^2\,|\,n\right\rangle\right\rangle$ under the integral leads to:

$$\frac{\partial^2}{\partial\mu\partial\nu}\delta(\mu-\nu)\quad. \tag{96}$$

Selecting eigenfunctions $\psi_{\mu,n}$ as real, one will obtain:

$$\left\langle n\,|\,u^2\,|\,n\right\rangle = \int d\mu\left[\left(\frac{\partial r_n}{\partial\mu}\right)^2 + r_n^2\left(\frac{\partial\varphi_n}{\partial\mu}\right)^2\right] + \int d\mu\, r_n^2\int dx\left(\frac{\partial}{\partial\mu}\psi_{\mu,n}\right)^2 -$$
$$-\frac{2t}{h}\int d\mu\, r_n^2\frac{\partial\varphi_n}{\partial\mu}V_{\mu,n} + \frac{t^2}{h^2}\int d\mu\, r_n^2 V_{\mu,n}^2. \tag{97}$$

**19.3.** Let us note that for the weight functions $\rho_n(\mu)$ with a small dispersion of the parameter $\mu$ it is possible to approximate that:

$$\frac{t^2}{\eta^2}\left\langle\left\langle V_{\mu,n}^2\right\rangle\right\rangle \approx \left(\frac{t}{\eta}\left\langle\left\langle V_{\mu,n}\right\rangle\right\rangle\right)^2\quad. \tag{98}$$

This means that the $\mu$ –packets do not segregate with time only for the distributions $\rho_n(\mu)$ as narrow peaks. However, if changing $\rho_n^2(\mu)$ by distribution with the $\delta(\mu-1)$ the constant part of the dispersion of the field phase $D_u$ becomes infinite.

# XX. "AGING" OF PARTIAL WAVE FUNCTIONS

The dynamics of changing of contributions by the future and by the past in partial wave function of stationary state can be monitored. To do that the integral which characterizes the contribution by the past may be used:

$$I_p(t) = \int\limits_0^\infty du \int dx F_n^*(u,x,t) F_n(u,x,t) \tag{99}$$

For normalized μ-packet

$$F_n(u,x,t) = \frac{1}{\sqrt{2\pi}} \int d\mu r_n(\mu) \psi_{\mu,n}(x) \exp\{i\varphi_n(\mu) - i\mu(\hbar^{-1} E_{\mu,n} t + u)\}, \tag{100}$$

$\varphi_n = \mu \hbar^{-1} t_{0n} \int d\mu r_n^2 V_{\mu,n}$, where $t_{0n}$ is a time point when function $\langle\langle n \mid u \mid n \rangle\rangle$ changes sign, can be chosen. According to definition (99), it will be:

$$I_p(t) = \frac{1}{2} + \frac{i}{2\pi} \int\limits_0^\infty du \int dx \int d\mu \int d\nu r_n(\mu) r_n(\nu) \psi_{\mu n}^*(x) \psi_{\nu n}(x) \sin(u(\mu-\nu))$$
$$\exp\{-i\varphi(\mu) + i\varphi(\nu) + i\hbar^{-1} t(\mu E_{\mu n} - \nu E_{\nu n})\}. \tag{101}$$

Formula (101) describes S-shaped time function with a bend-point at the moment $t_{0n}$. We are interested in the speed of change of the function at the bend-point. It is sufficient, considering $t_{0n} = 0$, to calculate the function $I_p(t)$ in linear approximation $I_p(t) = \frac{1}{2} + Ct$. Characteristic time period during which $I_p(t)$ changes from $I_p = 0$ to $I_p = 1$, is estimated as $T_n = C^{-1}$. Later we will consider $T_n$ estimation for the time of quantum transition as a result of which the $n$ state emerges.

In order to calculate the value of $C$ let us use the balance equation. According to (24),

$$C = \frac{\partial}{\partial t} \int\limits_0^\infty du \int dx F_n^*(u,x,t) F_n(u,x,t)|_{t=0} = \frac{1}{\hbar} \int dx F_n^*(0,x,0) V(x) F_n(0,x,0). \tag{102}$$

In equation (102) we consider the contribution, which corresponds to the limit $u \to \infty$, as a zero contribution, referring to an applicability of Riemann lemma. Let us write the value of $C$ in an explicit manner and consider the formula (76).

$$C = -\frac{1}{\text{h}} \int dx \int d\mu \int d\nu \, r_n(\mu) r_n(\nu) \psi^*_{\mu,n}(x) \psi_{\nu,n}(x) V(x) =$$

$$= -\frac{1}{\text{h}} \int d\mu \int d\nu \, r_n(\mu) r_n(\nu) \frac{\mu E_{\mu,n} - \nu E_{\nu,n}}{\mu - \nu} \langle \mu, n | \nu, n \rangle.$$

(103)

In the further estimations we will have to consider that the weight function corresponds to a rather narrow normalized peak $\left( |\mu - \nu| \ll 1 \right)$. In addition, we can consider that

$$\frac{\mu E_{\mu,n} - \nu E_{\nu,n}}{\mu - \nu} = V_n(\mu),$$

(104)

$$\langle \mu, n | \nu, n \rangle = 1.$$

(105)

Under such conditions we will have

$$C = -\frac{1}{\text{h}} \int d\mu \, r_n^2(\mu) V_{\mu,n},$$

(106)

$$T_n = -\text{h} / \int d\mu \, r_n^2(\mu) V_{\mu,n}.$$

(107)

Considering (107) a phase formula (95) can be written as

$$<< n \left| u \right| n >> = (t - t_{0n}) / T_n. \qquad (108)$$

One can see that arbitrary choosing the counting off level for the $V_{\mu,n}$ and the type of the weight function, it is possible to get the value of $T_n$, which would correspond to experimental data.

# XXI. ACCOUNTING FOR OF THE MOMENTS OF CREATION OF THE QUANTUM STATES

     Schemes for calculating probabilities for measurement outcomes, described in the chapter 10, do not consider a possibility that the occurrence of the quantum transition at the moment $t$, following the measurement, may be connected to the moment $t - \tau$ of the previous quantum transition during which the disturbed state has occurred. Indeed, the wave function, calculated as a solution of Schrödinger equation, does not contain information about the time point, at which the quantum transition into an initial state that is being evaluated, has occurred. However, the structure of the partial wave functions contains an indication of the moment, from which almost all paths coming from the past contribute to forming the partial function. Exactly this moment, $t - \tau$, may be considered a moment of creation of the initial quantum state. Quantum transition at the moment $t$ is also accompanied by the creation of a new state. In the current analysis we will just consider time intervals of these quantum transitions equal $T_s$.

     Let us take a look at the transitions between energy levels caused by a non-stationary potential. Let index s correspond to an initial stationary state of the system. A bilinear variant convenient for analyzing temporal structure of the quantum transition should be built. A bilinear variant, which is substantial, but not obligatory positive, satisfying above-mentioned conditions, is a symmetrized invariant $S$, composed for partial functions with selected moments of "creation" $t$ and $t - \tau$. Information about transitory moments is carried by the phases $\varphi_0 = -\mu(t-\tau)/T_s$ and $\varphi_1 = -\mu t / T_s$, as it follows from (94) and (107). As a result, instead of an original solution for non-stationary problem

$$F_0(u,x,t) = \frac{1}{\sqrt{2\pi}} \int d\mu r(\mu) \psi_\mu(x,t) \exp(i\varphi_0 - i\mu u), \qquad (109)$$

two new partial functions will be obtained

$$F_1(u,x,t) = \frac{1}{\sqrt{2\pi}} \int d\mu r(\mu) \psi_\mu(x,t) \exp\{i\varphi_0 - \mu(t-\tau)/T_s - i\mu u\},$$ (110)

$$F_2(u,x,t) = \frac{1}{\sqrt{2\pi}} \int d\mu r(\mu) \psi_\mu(x,t) \exp\{i\varphi_0 - \mu t/T_s - i\mu u\}.$$ (111)

According to the results of chapter 7 these functions are solutions of the initial problem with    shifted level for counting off the potential. Calculating invariant (33) gives us

$$S_{12} = \int d\mu r^2(\mu) \cos(\mu\tau/T_s).$$ (112)

It remains to take into account that the calculation results do not depend on the weight functions choice in case of $r^2(\mu) \rightarrow \delta(\mu-1)$.

Finally, we have:

$$S_{12} = \cos(\tau/T_s).$$ (113)

If, at the beginning of the calculation, to project the function $\psi_\mu(x,t)$ according to orthonormalized functions of the physical value $a_{\mu,n}(x)$, we will obtain in case of the narrow peak type weight function:

$$S_{12}(\tau) = \cos(\tau/T_s) \sum_n \left| < a_n(x) | \psi(x,t) > \right|^2.$$ (114)

For every $\tau$ value contributions of $W_n$ to this projection are analogous to the probabilities for the measurement outcomes $P_n$:

$$W_n = \cos(\tau/T_s) \left| < a_n(x) | \psi(x,t) > \right|^2.$$ (115)

However such sign changing expression can not be interpreted as distribution of the probabilities of quantum transitions; only those values of $\tau$ for which $W_n < 0$ can be interpreted as a condition of an absence of quantum transition. Under such interpretation, scale of the intervals $\tau$ becomes broken into alternating regions of $\pi T_s$ length that are "favorable" and "unfavorable" for quantum transitions. Conducting special experiments, in which the moments $t$ and $t-\tau$ are being registered, one may hope to find "quantum windows" $\pi/2 + 2n\pi \leq \tau \leq 3\pi/2 + 2n\pi$, $(n=0,1,2,..)$ that "prohibited" values of $\tau$ fall into. The function $W(\tau)$ that equals to 1 at permitted values of $\tau$ and equals to zero at "prohibited" values of $\tau$ we call as index of transition possibility.

## XXII. THE SYSTEM OF MANY MASSIVE PARTICLES

The formulas obtained in previous chapters are correct for the system of many massive spinless particles. In such case $\hat{T}$ is an Hermitian operator, which is equal to the sum of kinetic energy operators of the particles in the system. Potential $V(x,t)$ is a function, which depends on coordinates of all the particles included into the system. Generalized virial $u$, as well as time, is a general parameter for all particles in the system. For simplicity, let us analyze a system of two non-interacting localized particles. The particles (let us call them «1» and «2»), are in a state with energies $E_{1,i}$ and $E_{2,k}$, the states can change independently in case of quantum transitions (we are considering energy transitions only for simplicity' sake). The full energy of the system $E_{i,k} = E_{1,i} + E_{2,k}$ and it changes at the moment when at least one of the particles experiences quantum transition regarding energy. The last time transition moment of at least one particle in an analyzed coherent system should be considered a transition moment for any particle included in the system. If there is a time count off between consequent quantum transitions for a particle 1, then this count off should be conducted from the latest transition in the system, even if the particle that underwent transition is not a particle 1. Thus, the presence of many particles described by the single wave partial function in the system, even if these particles do not interact, may influence the picture of quantum transition.

This conclusion relates to the hypothetical so-called counterfactual situations in quantum mechanics discussed, for example, in [12]. The most important aspect of this conclusion is that it shows that under normal conditions in experiments analyzing distribution of $\tau$ value, the most of $\tau$ values coincide with time the application of disturbances.

## XXIII. CONCLUSION

**1a.** Reading in this paper the numerous considerations about the contributions of the "past" and the "future" on the PWF the reader may think "How contribution of the future can be discussed at all if any arbitrary change of the potential V in the future time interval changes the PWF at the present moment of time?". Certainly, in this paper the form of the potential is subjected to the strict constrains that allow us to write PWF in the definite form. Let us examine these constrains.

Majority of the significant results of this work were obtained by means of PWF as μ-packets with the potentials constant in time. In standard quantum mechanics stationary solutions can be used any time when there is a time interval, on which the potential is stationary. In case of μ-packets, stricter constrains are applied: the potential is considered to be constant on the whole interval $-\infty < t < \infty$. One may say that in this case the condition of the absolutely definite past and future are subjected. In this case is no place for doubts expressed at beginning of this chapter.

**1b.** The formulas for non-stationary DWE solutions are also present in this work. One should not forget that non-stationary solutions are used only for calculation of transition probabilities out of the fixed states. The probabilities of transitions between the states with different energy levels were examined earlier. Non-stationary solution of DWE in such case was shown in the chapter 16.3, formulas (73-35). These formulas have been designed in such way that they contain only the values of the potential, known till the moment $t$ we are interested in. This is the basis for applicability of the causality principle to solving non-stationary DWE.

**2.** The formal method of the time interval $T_n$ attribution to the process of creation of the state $n$ is described above. The moment of creation is identical to the moment of the change of $<<u>>$ sign for $\mu$-packets with $n$ index (the moment of balance between the past and the future). The $T_n$ value according to equation (108) is the intrinsic characteristics of the system state. The mean value of the field virial of the state is the quantity of periods $T_n$ after the moment of $n$ state creation. By this formalism one can obtain the connection between the measurable values $\tau$ and $T_n$ that not depend on the choice of $\mu$-packet. Nevertheless the initial description in the form of continual in time PWF means that the process of creation, in principle, is infinite in time.

**3.** The interpretation of the time structure of the interval between the consequent quantum transitions in the present work corresponds to the simplest rigid scheme of transitions impossibility for which the analogue of probability is negative. It is no dependence of the mode of this value. The application of more sophisticated analysis in this case can result in some differences in the time interval structure description.

# XIV. FIGURES

**Figures captures**

Fig.1 (a,b,c). Schematics of the relative distribution of past and future contributions during the process of state evolution; (1a) – advantage of the contribution of the states from the future, (1b) equality of contributions of the past and future, (1c) - advantage of contributions of the states of the past.

Fig 2. Schematics of the time of the quantum transition $T_n$ calculation (creation of the state in the form of $\mu$-packets) using the index of past $I_p$.

Fig. 3 (a,b). Schematics of $f(\tau)$ distribution, index of transition possibility $W(\tau)$, $f_0$ – distribution without $W(\tau)$ – index of transition possibility; (3a) – distribution if $<\tau> << T$; (3b) - $f(\tau)$ distribution if $<\tau> >> T$.

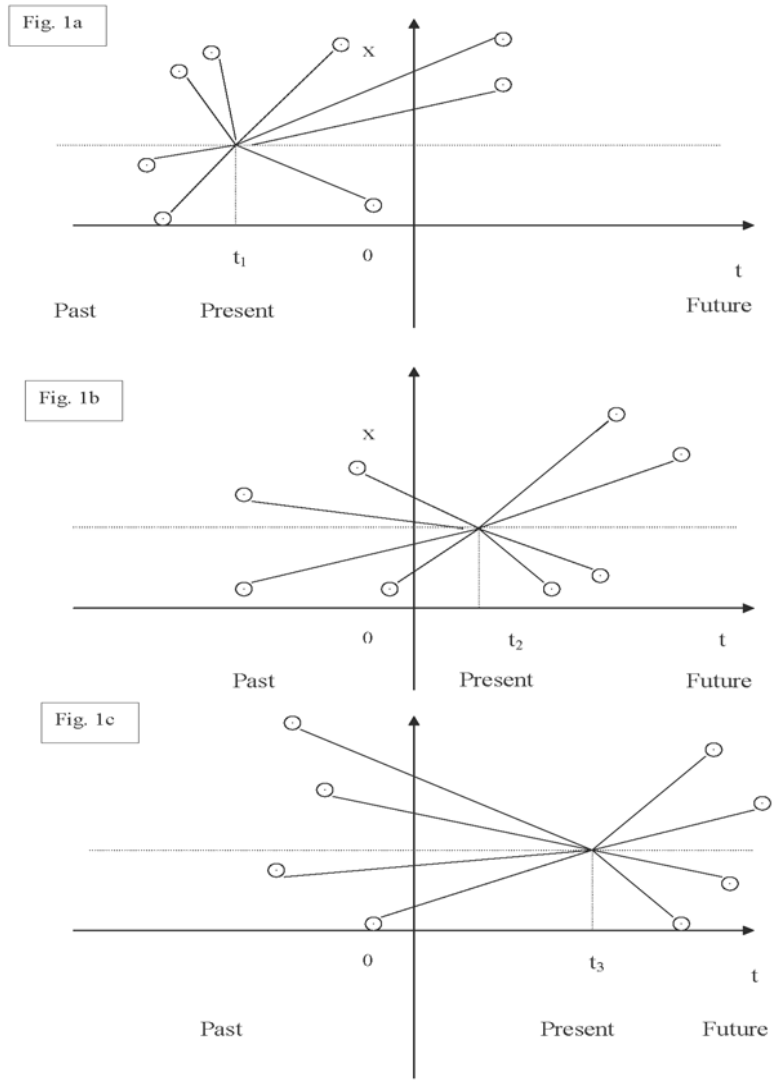

Fig.1

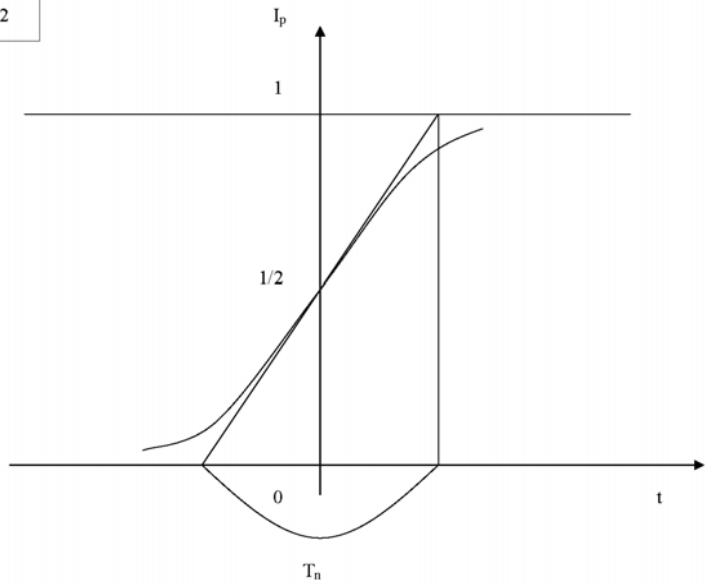

Fig. 2

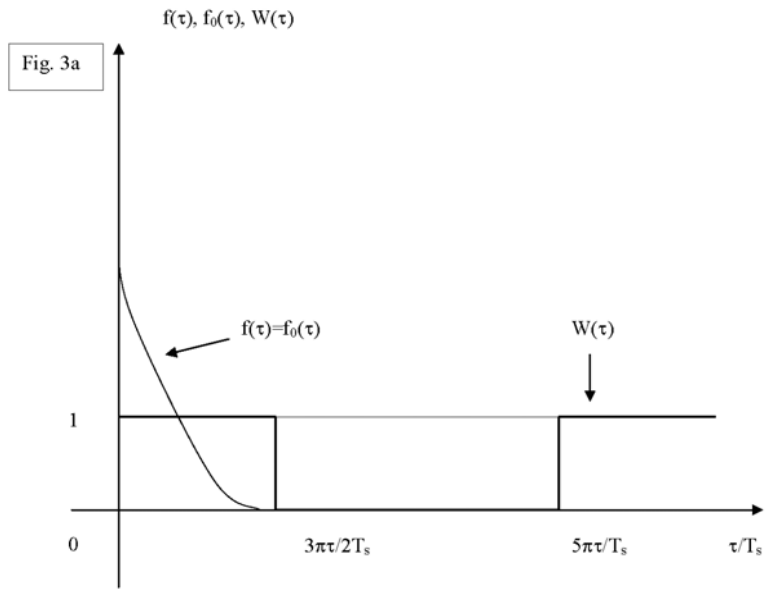

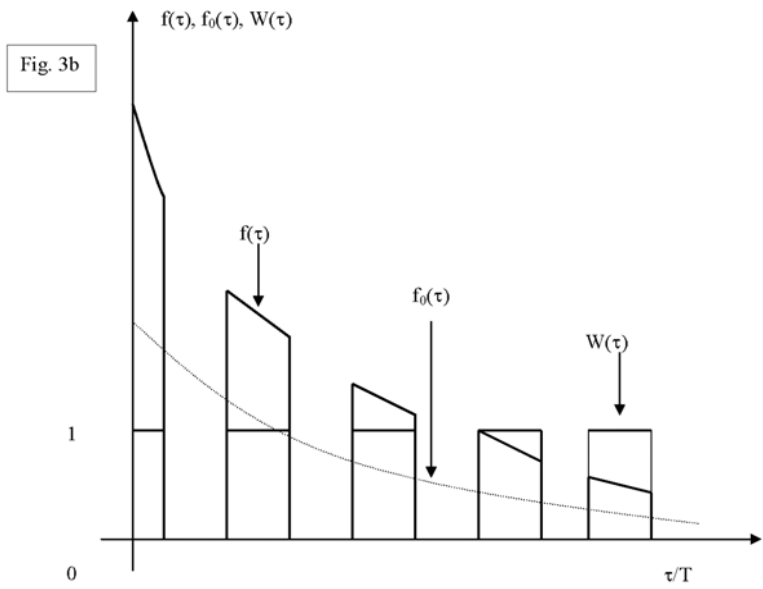

Fig. 3

**Acknowledgements**

The author expresses gratitude to D. B. Kucher, K. A. Tolstoluzhsky, and L. A. Pastur for the interest in this investigation at its early stage. The author is also grateful to A. O. Rudenko, T. Z. Sarkisian and Y. G. Shckorbatov for invaluable assistance in the process of preparing this publication.

# APPENDIX A.

## PARAMETRIZING MOVEMENTS OF CLASSIC PARTICLE BY MEANS OF THE FIELD VIRIAL

The field virial $u$ can be considered an analog of time $t$ in quantum mechanics. The value of the integral (7) can also be used for path parameterization in classical dynamics problems. However, in that case, one has to consider a one-dimensional movement only. Let $x$ be the coordinate of dot on the straight line, along which the particle with mass $m$ moves. If potential energy is $V[x]$ the equation of this movement will be:

$$m\frac{d^2x}{dt^2} = -\frac{dV(x)}{dx}.$$ (A.1)

Let us parametrize a path of the particle by means of new parameter $\tau$, which has the dimension of action and is connected to the time $t$ by

$$d\tau = V(x(t))dt \quad \text{or} \quad \tau(t) = \int_{t_1}^{t} dt' V(x(t))$$ (A.2)

It is also convenient to exchange coordinate $x$ for parameter $z$ according to the rule:

$$dz = V(x)dx \quad \text{or} \quad z(x) = \int_{x_1}^{x} dx' V(x').$$ (A.3)

In case if potential *V[x]* has a constant sign, formulas (A.2) and (A.3) set reciprocal relation between values $t$ and $\tau$, $x$ and $z$. Having changed variables in (A.1) and considering potential energy as a function $\widetilde{V}(z)$, we will obtain an equation of movement:

$$m\frac{d^2z}{d\tau^2} = -\frac{d\widetilde{V}(z)}{dz}.$$

(A.4)

The dimensionless value $u$ has been preferentially used instead of $\tau$ in this work.

# APPENDIX B.

# THEOREM ABOUT THE AVERAGE VALUE OF THE POTENTIAL

**B.1.** Let us prove the following theorem. Consider eigenfunctions $\psi_{\lambda,l}(x)$ and $\psi_{\nu,n}(x)$ from the stationary Schrödinger's equation (38) that have effective mass $\mu m$ correspond to two values $\mu$ ($\mu_1 = \lambda$ and $\mu_2 = \nu$) and two energy levels with numbers l and n. Then the following equality is correct:

$$(\lambda - \nu)\langle \lambda, l \mid V \mid \nu, n \rangle = (\lambda E_{\lambda,l} - \nu E_{\nu,n})\langle \lambda, l \mid \nu, n \rangle$$

(B.1)

here

$$\langle \lambda, l \mid \nu, n \rangle = \int dx\, \psi_{\lambda,l}^*(x)\psi_{\nu,n}(x),$$

(B.2)

$$\langle \lambda, l \mid V \mid \nu, n \rangle = \int dx\, \psi_{\lambda,l}^*(x)V(x)\psi_{\nu,n}(x).$$

(B.3)

In order to prove this theorem, let us write out the stationary Schrödinger's equations with the parameter $\nu$ and the parameter $\lambda$ (in the following $\hat{T}$ – kinetic energy operator):

$$\overset{l}{T}\psi_{\nu,n}(x) = \nu E_{\nu,n}\psi_{\nu,n}(x) - \nu V(x)\psi_{\nu,n}(x) \ , \tag{B.4}$$

$$\overset{l}{T}\psi_{\lambda,l}(x) = \lambda E_{\lambda,l}\psi_{\lambda,l}(x) - \lambda V(x)\psi_{\lambda,l}(x). \tag{B.5}$$

Let us multiply the equation (B.4) by the $\psi^*_{\lambda,l}$, and the equation (B.5) by the $\psi^*_{\nu,n}$, and then integrate by $x$. Then we will apply an operation of complex correspondence to the first of the equalities obtained. As operators $\overset{l}{T}$ and $\overset{l}{V}$ are Hermitian operators, then $\left\langle \lambda,l \,|\, \overset{l}{T} \,|\, \nu,n \right\rangle^* = \left\langle \nu,n \,|\, \overset{l}{T} \,|\, \lambda,l \right\rangle$ and $\left\langle \lambda,l \,|\, V \,|\, \nu,n \right\rangle^* = \left\langle \nu,n \,|\, V \,|\, \lambda,l \right\rangle$. In addition, $\left\langle \nu,n \,|\, \lambda,l \right\rangle^* = \left\langle \lambda,l \,|\, \nu,n \right\rangle$. From these equalities it follows:

$$\nu E_{\nu,n}\langle \nu,n \,|\, \lambda,l \rangle - \nu\langle \nu,n \,|\, V \,|\, \lambda,l \rangle = \lambda E_{\lambda,l}\langle \nu,n \,|\, \lambda,l \rangle - \lambda\langle \nu,n \,|\, V \,|\, \lambda,l \rangle. \tag{B.6}$$

This equality is analogous to a (B.1).

**B.3.** Let us return to the condition $\lambda E_{\lambda,l} = \nu E_{\nu,n}$ (52) and equality $\langle \lambda,l \,|\, V \,|\, \nu,n \rangle = 0$ (53), and elucidate whether these equalities can be executed if n=l. In the later case there should be such values $\mu_1$ and $\mu_2$ for which:

$$\mu_1 E_{\mu_1,n} = \mu_2 E_{\mu_2,n} \tag{B.7}$$

If to choose the level of energy count off for the localized state in such way that $E_{\mu,n} < 0$, then

$$\frac{\partial}{\partial\mu}\big(\mu E_{\mu,n}\big) = E_{\mu,n} + \mu\frac{\partial}{\partial\mu}E_{\mu,n} < 0 \ . \tag{B.8}$$

As a result, $\mu E_{\mu,n}$ becomes a monotonous $\mu$ function and the equality (B.7) may take place if $\mu_1 = \mu_2$ only.

**B.4.** Differentiating in equality (B.1) by $\lambda$ and then considering $\lambda = \nu$, a useful equation can be obtained:

$$\int dx \psi_{\lambda,l}^{*}(x) \frac{\partial}{\partial \lambda} \psi_{\lambda,n}(x) = \left\langle \lambda, l | V | \lambda, n \right\rangle \left[ \lambda \left( E_{\lambda,n} - E_{\lambda,l} \right) \right]^{-1}. \qquad (B.9)$$

Differentiating in the equality (B.1) by $\lambda$ and by $\nu$ with n=l, and then letting $\lambda = \nu$, a number of other consequences can be obtained, considering that functions $\psi$ are real and numerated by 1. Let us use:

.

$$\frac{\partial}{\partial \mu} \psi_{\mu,n} = \psi_{\mu,n}^{/}, \quad \frac{\partial^{2}}{\partial \mu^{2}} \psi_{\mu,n} = \psi_{\mu,n}^{//} \qquad (B.10)$$

Then:

$$\left\langle \psi_{\mu,n}^{\prime} | V | \psi_{\mu,n}^{\prime} \right\rangle = \frac{\partial}{\partial \mu} (\mu E_{\mu,n}) \left\langle \psi_{\mu,n}^{\prime} | \psi_{\mu,n}^{\prime} \right\rangle + \frac{1}{6} \frac{\partial^{3}}{\partial \mu^{3}} (\mu E_{\mu,n}), \qquad (B.11)$$

$$2 \left\langle \psi_{\mu,n}^{\prime} | V | \psi_{\mu,n}^{\prime} \right\rangle - \left\langle \psi_{\mu,n}^{\prime\prime} | V | \psi_{\mu,n} \right\rangle = 3 \frac{\partial}{\partial \mu} (\mu E_{\mu,n}) \left\langle \psi_{\mu,n}^{\prime} | \psi_{\mu,n}^{\prime} \right\rangle. \qquad (B.12)$$

# APPENDIX C.

# INVARIANT IN THE SPACE $\{t, x\}$ WITH STATIONARY POTENTIAL

Let us observe the evolution of partial wave functions $F(u, x, t)$, using the field phase $u$ as an analogue of time; actual time $t$ will be considered an additional variable, by which integration should be done normalizing $F(u, x, t)$. Solution of DWE (9) has been developed above as a resolution by orthonormalized μ–packets (43). Therefore, it should be sufficient for us to test preservation of $M_{n,l}$ value for the pair of μ-packets with arbitrary indexes $n$ and $l$:

$$M_{n,l}(u) = \frac{1}{\eta} \int dx \int dt\, V(x) F_n^*(u,x,t) F_l(u,x,t)\,. \tag{C.1}$$

Using definitions (25) for $F_n$ and $F_l$, it is convenient instead of integrating parameters $\nu$ and $\lambda$, to switch to new parameters $p$ and $q$ that have an energy dimension:

$$p(\nu) = \nu E_{\nu,n}, \quad q(\lambda) = \lambda E_{\lambda,l} \tag{C.2}$$

It can be shown (see Supplement 2) that it is possible to develop reciprocal relation $p = f_n(\nu)$, $q = \varphi_l(\lambda)$ for localized states with indexes $n$ and $l$. In this case it follows from the equation (C.2) that

$$\frac{\partial f_n(\nu)}{\partial \nu} = V_{\nu,n}, \qquad \frac{\partial \varphi_l(\lambda)}{\partial \lambda} = V_{\lambda,l} \tag{C.3}$$

In the following calculations we will assume that $V_{\nu,n}$ and $V_{\lambda,l}$ retain the same sign in the spaces of values $\nu$ and $\lambda$, that will be designated as $\operatorname{sgn}(V_{\nu,n}, V_{\lambda,l})$. One should note, that this condition is less restricting then constant sign of the function $V(x)$. Integral $M_{n,l}(u)$ can be written as:

$$M_{n,l}(u) = \frac{1}{2\pi\hbar} \int dx \int dt \int_{p_-}^{p_+} dp \int_{q_-}^{q_+} dq\, \frac{\tilde{\rho}_n^*(p)}{l_{p,n}^{\frac{2}{3}}} \frac{\tilde{\rho}_l(q)}{l_{q,l}^{\frac{2}{3}}}$$
$$\exp\left[iu\left(\frac{p}{\tilde{E}_{p,n}} - \frac{q}{\tilde{E}_{q,l}}\right)\right] \tilde{\psi}_{p,n}^*(x) \tilde{\psi}_{q,l}(x) V(x) \exp\left[\frac{i}{\hbar}(p-q)t\right] \tag{C.4}$$

The functions $\tilde{\rho}_n(p) = \rho_n(\nu(p))$, $\tilde{\psi}_{p,n}(x) = \psi_{\nu(p),n}(x)$, $\tilde{E}_{p,n} = E_{\nu(p),n}$ and $\tilde{V}_{p,n} = V_{\nu(p),n}$. are introduced here. Analogous designations are introduced for parameter $q$ and index $l$. Upon integrating by $t$, $\eta\delta(p-q)$ is obtained in under integral expression. As a result:

$$M_{n,l} = \operatorname{sgn}(V_{\nu,n}, V_{\lambda,l}) \int_{p_-}^{p_+} dp\, \frac{\tilde{\rho}_n^*(p)\tilde{\rho}_l(p)}{l_{p,n}^{\frac{2}{3}} l_{p,l}^{\frac{2}{3}}} \exp\left[iup\left(\tilde{E}_{p,n}^{-1} - \tilde{E}_{p,l}^{-1}\right)\right] \langle \tilde{\psi}_{p,n} | V | \tilde{\psi}_{p,l}\rangle\,. \tag{C.5}$$

Let us consider n=l. In this case under integral expression differs from zero if $p = \nu(p)E_{\nu(p),n} = \lambda(p)E_{\lambda(p),n}$, which leads to the conditions $\nu(p) = \lambda(p)$ and $E_{\nu(p),n} = E_{\lambda(p),n}$, (see. equation (S2.11)). Taking into account the last equality leads to an independence of $M_{n,l}$ from $u$. Considering the integral over parameter $\nu$ and assuming $dp = V_{\nu,n}d\nu$ :

$$M_{n,n} = \text{sgn}(V_{\nu,n})\int_{\nu_-}^{\nu} d\nu \frac{|\rho_n(\nu)|^2}{V_{\nu,n}} \langle \psi_{\nu,n} \mid V \mid \psi_{\nu,n} \rangle \tag{C.6}$$

According to (49), $V_{\nu,n} = \langle \psi_{\nu,n} \mid V \mid \psi_{\nu,n} \rangle$, therefore

$$M_{n,n} = \text{sgn}(V_{\nu,n})\int_{\nu_-}^{\nu_+} d\nu |\rho_n(\nu)|^2 . \tag{C.7}$$

For the orthonormalized $\mu$-packets

$$M_{n,n} = \text{sgn}(V_{\nu,n}) \tag{C.8}$$

It remains to analyze a case when $n \neq l$. The integrals included in $M_{n,l}$ can be viewed as:

$$\langle \widetilde{\psi}_{p,n} \mid V \mid \widetilde{\psi}_{p,l} \rangle = \langle \psi_{\nu(p),n} \mid V \mid \psi_{\lambda(p),l} \rangle \tag{C.9}$$

If $n \neq l$ under integral expression in $M_{n,l}$ differs from zero under condition $p = \nu(p)E_{\nu(p),n} = \lambda(p)E_{\lambda(p),l}$. If the equality $\nu(p) = \lambda(p) = \varphi$ is correct, the above-mentioned condition is reduced to $E_{\varphi,n} = E_{\varphi,l}$, i.e. to coinciding of energy values with different indexes, which is impossible. Therefore, one has to assume that $\nu(p) \neq \lambda(p)$. However, according to (S2.1), it should be $\langle \nu,n \mid V \mid \lambda,l \rangle = 0$. Hence, $M_{n,l} = 0$, which means the orthogonality of the $\mu$-packets in an space $\{t,x\}$ with weight $V(x)$.

$$M_{n,l} = \text{sgn}(V_{\nu,n},V_{\lambda,l})\delta_{n,l} . \tag{C.10}$$

If function $F(u,x,t)$ is built as a linear combination of $\mu$-packets, integrating by $x$ and by $t$, one will obtain:

$$M = \text{sgn}(V)\sum_i |b_i|^2 \tag{C.!1}$$

To normalize a linear combination with $\mathrm{sgn}(V)$ independent from the μ-packet index, if is sufficient to use $\mathrm{sgn}(V)V(x)$ as a weight function and to require fulfillment of the condition (66).

Using this result, it is also possible to use invariant (C.1) for the problems, in which non-stationarity of the potential occurs due to a small addition to a stationary potential, and this addition differs from zero during a finite time interval only.

# APPENDIX D.

# CALCULATION OF THE VALUE $T$ MOMENTS IN THE SPACE $\{t,x\}$ UNDER STATIONARY POTENTIAL.

Calculation $\left\langle \left\{ t^{k} \right\} \right\rangle$ for the μ-packets normalized in the $\{t,x\}$ space leads

$$
\begin{aligned}
\left\langle \left\{ n \mid t \mid n \right\} \right\rangle = & -\hbar \,\mathrm{Im} \int_{\mu_{-}}^{\mu_{+}} d\mu\, \rho_n\left(\mu\right) \frac{\partial}{\partial \mu}\left(\rho_n^{*}\left(\mu\right) V_{\mu,n}^{-1}\right) - \\
& -\hbar \,\mathrm{Im} \int_{\mu_{-}}^{\mu_{+}} d\mu\, \left|\rho_n\left(\mu\right)\right|^2 V_{\mu,n}^{-2} \int dx V(x) \frac{\partial \psi_{\mu,n}^{*}(x)}{\partial \mu} \psi_{\mu,n}(x) - \\
& -u\hbar \int_{\mu_{-}}^{\mu_{+}} d\mu\, \left|\rho_n\left(\mu\right)\right|^2 \quad V_{\mu,n}^{-1}.
\end{aligned}
\tag{D1}
$$

$$
\begin{aligned}
\left\{ n \mid t^2 \mid n \right\} = & \,\hbar^2 \int d\mu \left|\frac{\partial}{\partial \mu}\left(\rho_n\left(\mu\right) V_{\mu,n}^{-1}\right)\right|^2 + \\
& +\hbar^2 \int d\mu \left|\rho_n\left(\mu\right)\right|^2 V_{\mu,n}^{-3} \int dx V(x) \left|\frac{\partial}{\partial \mu}\psi_{\mu,n}(x)\right|^2 + \\
& +2\hbar^2 \,\mathrm{Re} \int d\mu\, \rho_n\left(\mu\right) V_{\mu,n}^{-2} \frac{\partial}{\partial \mu}\left(\rho_n^{*}(\mu) V_{\mu,n}^{-1}\right) \int dx V(x)\psi_{\mu,n}^{*}(x)\frac{\partial}{\partial \mu}\psi_{\mu,n}(x) + \\
& +2u\hbar^2 \,\mathrm{Im} \int d\mu \left|\rho_n\left(\mu\right)\right|^2 V_{\mu,n}^{-3} \int dx V(x)\psi_{\mu,n}(x)\frac{\partial}{\partial \mu}\psi_{\mu,n}^{*}(x) + \\
& +2u\hbar^2 \,\mathrm{Im} \int d\mu\, \rho_n(\mu) V_{\mu,n}^{-1} \frac{\partial}{\partial \mu}\left(\rho_n^{*}(\mu) V_{\mu,n}^{-1}\right) + \\
& +\hbar^2 u^2 \int d\mu \left|\rho_n\left(\mu\right)\right|^2 V_{\mu,n}^{-2}.
\end{aligned}
\tag{D2}
$$

As in case of fixed time $t$, it follows from (D1), (D2) that for a distribution $\rho_n(\mu)$ with a small dispersion of the parameter $g$ $\mu$-packets spread slowly, although in this case a constant part of dispersion for time intervals is large. If functions $\psi_{\mu,n}$ and $\rho_n(\mu)$ are chosen as real, the results are simplified:

$$\left\langle \{n\,|\,t\,|\,n\} \right\rangle = -u\hbar\int d\mu \rho_n(\mu)^2 V_{\mu,n}^{-1} \tag{D3}$$

$$\left\langle \{n\,|\,t^2\,|\,n\} \right\rangle = \hbar^2 \int d\mu\, V_{\mu,n}^{-1} \frac{\partial}{\partial \mu}\left(\rho_n(\mu)V_{\mu,n}^{-1}\right)\frac{\partial}{\partial \mu}\rho_n(\mu) +$$
$$+ \hbar^2 \int d\mu \rho_n^2(\mu) V_{\mu,n}^{-3} \int dx V(x)\left(\frac{\partial}{\partial \mu}\psi_{\mu,n}(x)\right)^2 + \tag{D4}$$
$$+ \hbar^2 u^2 \int d\mu \left(\rho_n(\mu)V_{\mu,n}^{-1}\right)^2$$